\documentclass[fleqn,12pt]{wlscirep}
\usepackage[font=small,labelfont=bf]{caption}
\usepackage[right]{lineno}
\usepackage{amsmath,mleftright}

\usepackage[backend=biber,sorting=none,doi=false,isbn=false,url=false,eprint=false]{biblatex}
\title{A survey of open questions in adaptive therapy: bridging mathematics and clinical translation}
\author[1,*]{Jeffrey West}
\author[3,4]{Fred Adler}
\author[1]{Jill Gallaher}
\author[1]{Maximilian Strobl}
\author[1]{Renee Brady-Nicholls}
\author[1]{Joel S. Brown}
\author[1]{Mark Robertson-Tessi}
\author[6,*]{Eunjung Kim}
\author[5,*]{Robert Noble}
\author[2,*]{Yannick Viossat}
\author[1,*]{David Basanta}
\author[1,*]{Alexander R. A. Anderson}

\affil[1]{Department of Integrated Mathematical Oncology, H. Lee Moffitt Cancer Center \& Research Institute. Tampa, FL 33612, USA}
\affil[2]{Ceremade, Université Paris-Dauphine, Université Paris Sciences et Lettres, Paris, France}
\affil[3]{Department of Mathematics, University of Utah, Salt Lake City, UT 84112, USA}
\affil[4]{School of Biological Sciences, University of Utah, Salt Lake City, UT 84112, USA}
\affil[5]{Department of Mathematics, City, University of London, London, UK}
\affil[6]{Korea Institute of Science and Technology}

\affil[*]{{\bf Co-corresponding authors:}}
\affil[ ]{jeffrey.west@moffitt.org}
\affil[ ]{viossat@ceremade.dauphine.fr}
\affil[ ]{robert.noble@city.ac.uk}
\affil[ ]{eunjung.kim@kist.re.kr}
\affil[ ]{david@cancerevo.org}
\affil[ ]{alexander.anderson@moffitt.org}

\begin{document}
\maketitle

\subsection*{Abstract}
Adaptive therapy is a dynamic cancer treatment protocol that updates (or ``adapts'') treatment decisions in anticipation of evolving tumor dynamics. This broad term encompasses many possible dynamic treatment protocols of patient-specific dose modulation or dose timing. Adaptive therapy maintains high levels of tumor burden to benefit from the competitive suppression of treatment-sensitive subpopulations on treatment-resistant subpopulations. This evolution-based approach to cancer treatment has been integrated into several ongoing or planned clinical trials, including treatment of metastatic castrate resistant prostate cancer, ovarian cancer, and BRAF-mutant melanoma. In the previous few decades, experimental and clinical investigation of adaptive therapy has progressed synergistically with mathematical and computational modeling. In this work, we discuss 11 open questions in cancer adaptive therapy mathematical modeling. The questions are split into three sections: 1) the necessary components of mathematical models of adaptive therapy 2) design and validation of dosing protocols, and 3) challenges and opportunities in clinical translation.

\subsection*{Introduction: a survey of open questions in adaptive therapy}
\color{red}Jeffrey West, Eunjung Kim, Rob Noble, Yannick Viossat, David Basanta, Alexander Anderson: \color{black}

\noindent Treatment resistance in cancer therapy remains an overarching challenge across all types of cancer and all modes of treatment including targeted therapy, chemotherapy, and immunotherapy. Despite the ubiquity of the evolution of resistance, the ``more is better'' paradigm still prevails as standard of care. Over the past decade, a small group of oncologists in collaboration with evolutionary biologists and experimental biologists have proposed an ``adaptive therapy'' approach to cancer treatment~\cite{gatenby2009adaptive,  zhang2017integrating, zhang2022evolution}. Adaptive therapy maintains high levels of tumor burden in order to capitalize on competition between treatment-sensitive and treatment-resistant clones, and the potential cost of resistance.

\begin{figure}[t!]
    \centering
    \includegraphics[width=1\linewidth]{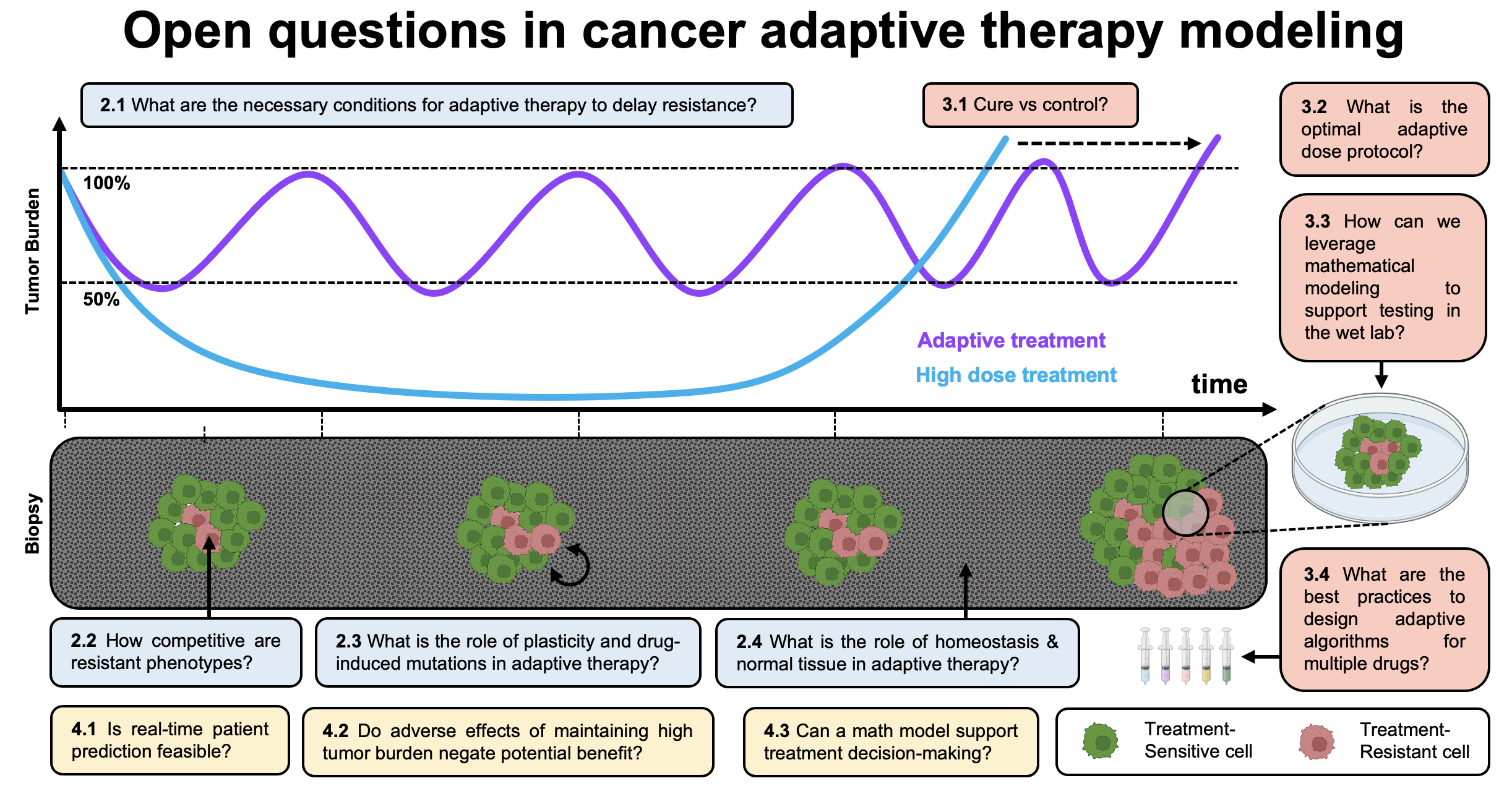}
    \caption{{\bf Open questions in adaptive cancer therapy modeling}: Schematic of tumor burden under maximum tolerable dose (blue) and adaptive dosing (purple), with corresponding biopsies. Adaptive therapy is designed to exploit competition between treatment-sensitive (green) and resistant (red) cells to prolong the emergence of resistance. 11 questions representing future challenges in the field of adaptive therapy are shown, and answered within the text. Questions are color-coded by section: the necessary components of mathematical models (blue), design and validation of dosing protocols (red), and challenges and opportunities in clinical translation (yellow).}
    \label{fig:adaptive_therapy_schematic}
\end{figure}

Adaptive therapy is characterized by dynamic treatment protocols which update (or ``adapt'') in anticipation to evolving tumor dynamics. These protocols are patient-specific, leading to variable dose modulation or dose timing between patients. While the term adaptive therapy is broad and encompasses many possible dynamic treatment protocols, the term is often used with specific reference to a recent pilot clinical trial in prostate cancer. This first adaptive trial enrolled a small cohort of metastatic castrate-resistant prostate cancer patients, contingent upon a minimum of 50\% drop in the level of prostate-specific antigen biomarker (PSA; a proxy for tumor burden) under abiraterone administration. Abiraterone is then withdrawn until PSA returns to pre-treatment levels and then restarted. This 50\% rule leads to treatment holidays that are patient-specific (treatment protocol varies considerably between patients). Holidays are often shorter in later treatment cycles when PSA dynamics speed up~\cite{zhang2017integrating}. Initial results of the trial indicate a prolonged progression-free survival and lower cumulative dose when compared to a contemporaneous cohort of patients receiving the standard of care~\cite{zhang2021response}. A schematic of adaptive therapy is shown in figure \ref{fig:adaptive_therapy_schematic} (purple), which prolongs relapse when compared to a high-dose schedule (blue). While initial results appear promising, this  trial was performed on a small cohort of men and did not include a randomized control arm~\cite{mistry2021reporting, zhang2021response}. A similar and larger, randomized trial in metastatic castrate-resistant prostate cancer is planned (ANZadapt; NCT05393791).

The first trial has created interest in designing new adaptive treatment protocols in prostate cancer as well as other cancers. Adaptive treatment protocols are often binned into two dose-scheduling approaches: dose modulation and dose skipping. Both are designed to prolong sensitivity to therapy and both have been tested experimentally~\cite{gatenby2009adaptive, enriquez2016exploiting}, while only dose skipping has been translated to the clinic~\cite{zhang2017integrating}. Throughout the text, we refer to the following treatment scheduling protocols:
\begin{enumerate}
    \item {\bf Maximum Tolerable Dose (MTD)}: periodic administration of a high dose, limited by toxic side effects.
    \item {\bf Intermittent Therapy}: periodic administration of a high dose with fixed, periodic treatment holidays.
    \item {\bf Adaptive Therapy (Dose Skipping)}: adaptive dosing where a high dose is administered until a desired tumor response (e.g. 50\% size reduction), followed by a treatment holiday until a desired upper threshold (e.g. 100\%) and repeated.
    \item {\bf Adaptive Therapy (Dose Modulation)}: adaptive dosing where dose is modulated (increased or decreased) at regular intervals depending on tumor response.
\end{enumerate}

In figure \ref{fig:adaptive_therapy_schematic}, we introduce eleven open questions regarding future directions of mathematical modeling in adaptive cancer therapy. These were the result of a four-day workshop on Cancer Adaptive Therapy Models (CATMo; https://catmo2020.org/) in December 2020. The conference brought together a multi-disciplinary group of mathematicians, clinical oncologists, and experimental biologists to discuss successes, challenges and opportunities in adaptive therapy. We have categorized these questions into three sections: 1) the necessary components of mathematical models, 2) the design and validation of dosing protocols, and 3) challenges and opportunities in clinical translation.

\section{The necessary components of mathematical models}
\subsection{What are the necessary conditions for adaptive therapy to delay resistance?}
\color{red}Fred Adler: \color{black} It is thought that the success of adaptive therapy in delaying the emergence of resistance depends on three characteristics of the cancer: a) resistance is costly, b) resistant cells can be suppressed by competition with sensitive cells, and c) therapy reduces the population of sensitive cells. Simple models based on these assumptions show that adaptive therapy can indeed delay the emergence of resistance. These simple models raise two further questions: 1) what are the appropriate objectives for evaluating the success of therapy? 2) do the main results hold up in models that include additional components of real tumors?

Our ultimate objective is to maximize survivorship or quality-of-life adjusted survivorship of the patient. This depends on the cancer burden, the treatment burden, and the effectiveness of treatment in suppressing the cancer in the long run~\cite{bayer2022markovian}.  In most cases, we do not have sufficient information to quantify each of these costs and benefits over the long run, but we can consider them in concert to evaluate overall success.

Models of adaptive therapy typically include distinct sensitive and resistant cancer cell populations, although some recent models follow a continuum of cell types~\cite{pressley2021evolutionary}. Model extensions include a) Healthy cells: these cells are always present within a tumor and they will interact with cancer cells~\cite{west2018capitalizing}, b) Immune cells: these cells can help control cancer but can themselves be affected by treatments~\cite{piretto2018combination, schattler2016dynamical, park2019goldilocks}, c) Resources: hormones~\cite{kareva2021estrogen} introduce delays and can alter evolutionary trajectories, and have been modeled as consumer-resource dynamics~\cite{zazoua2019analysis} and more mechanistic models with androgen dynamics~\cite{jain2011mathematical}. d) Allee effect: cell populations that grow more slowly (per capita) at low populations~\cite{konstorum2016feedback} effectively introduce an element of cooperation, e) Phenotypic plasticity: rapid changes in cell phenotypes can generate resistance far more quickly than mutation or population dynamics~\cite{salgia2018genetic}.

A key result from our work on the basic model is a tradeoff curve between time for resistant cells to emerge and the mean cancer burden~\cite{buhler2021mechanisms}.  This tradeoff holds for both adaptive (dose skipping) and intermittent therapies, and is robust across all model extensions except for the Allee effect and cell plasticity.  With an Allee effect, results are quite different. Aggressive therapy can drive cells below the threshold, and prevent both resistant cells and total cells from reaching their upper thresholds.  Adaptive therapy, by backing off early to avoid favoring resistance, can behave quite poorly, leading to escape times nearly as short as those with no therapy and with a high total cell burden.

With the exception of the success of high dose therapy with a strong Allee effect, no universal therapy can achieve all three objectives of lowering average dose, delaying time to emergence of resistant cells, and reducing total tumor burden.  All strategies show a tradeoff between delaying emergence of resistant cells and a high cancer burden. Choosing the appropriate treatment requires assessing the individual patients and specific cancers, and include factors often not included in models, such as therapy toxicity~\cite{ballesta2014physiologically}. Phenotypic plasticity, where resistance is induced by therapy rather than arising from mutations or pre-existing variants~\cite{salgia2018genetic}, makes resistance much more difficult to suppress~\cite{feizabadi2017modeling}.  Reversible behaviors can create complex responses to therapeutic timing~\cite{hirata2010development}.

Effective adaptive therapies require fitting data on individual patients, and data may lack the resolution to distinguish among alternative models. In the case of PSA in prostate cancer, a simple model~\cite{hirata2010development}, a more complex model with basic androgen dynamics~\cite{portz2012clinical}, and a detailed model of androgen dynamics~\cite{jain2011mathematical} all fit data on a set of patients reasonably well, although with some exceptions~\cite{hatano2015comparison}.  If models can be fit to the dynamics, adaptive therapies may be more robust to patient variability than prescribed timing of intermittent therapy. Although data may lack the resolution to identify specific mechanisms of interaction, such as the strength of competition between different cancer cell phenotypes, simple models may have the greatest potential to capture dynamics and guide therapy. The ideal combination will be patient-specific models combined with {\em in vivo} data, perhaps with mouse PDX models~\cite{siolas2013patient}, or {\em in vitro} data on patient derived cells that can reveal mechanisms of interaction in different treatment environments.

\subsection{How competitive are treatment resistant phenotypes?}
\color{red}Rob Noble: \color{black} Adaptive therapy aims to exploit competition between treatment-sensitive and resistant cells. Key questions remain largely unanswered. First, what is the nature of this competition? Mathematical modellers typically assume that the fitness of resistant cells is a simple function of their relative abundance and/or the total tumor size (reviewed in~\cite{viossat2021theoretical}). But frequency- or density-dependent mathematical functions only approximate average population dynamics. Actual clonal growth rates depend on the spatial arrangement of cells, their interaction ranges, and local levels of shared resources, all of which vary both within and between tumors~\cite{Noble2021_Modes, West2021_Normal, Fu2021}. For example, if a tumor grows mainly at its boundary then spatial constraints alone could suffice to contain rare resistant clones, but only if they are located away from the boundary~\cite{gallaher2018spatial, bacevic2017spatial}. A corollary is that the effectiveness of adaptive therapy may vary between cancer types due to different tumor architectures~\cite{Noble2021_Modes}. Although spatially-structured computational and experimental models can account for some important factors -- such as competition for space and oxygen -- the ability to predict clinical outcomes hinges on these models accurately matching the parameters of human intra-tumor cell-cell interactions, which remain largely uncharacterised. Further experimental studies and clinical image analyses are needed to quantify these parameters.

Second, are resistant cells less competitive? The seminal 2009 paper by Gatenby et. al.~\cite{gatenby2009adaptive} postulated that cells insensitive to therapy incur a fitness cost in the absence of treatment, which adaptive therapy can exploit. A reduction in cell proliferation rate or carrying capacity might result from cells diverting resources away from proliferation and towards breaking down or pumping out toxins. On the other hand, it is uncertain what fitness effects, if any, should result from mutations that modify specific drug targets. Experimental evidence is mixed. A study of tumor containment using a cyclin-dependent kinase inhibitor found a cost of resistance both {\em in vitro} and in mice~\cite{bacevic2017spatial}. Conversely, cancer cells resistant to the tyrosine kinase inhibitor alectinib have been observed outcompeting ancestral cells in co-culture~\cite{kaznatcheev2019fibroblasts}. Competition assays should in any case be interpreted with caution because there are many potential mechanisms of resistance to a given treatment and the relative fitness of each phenotype will vary with its microenvironment. Theoretical analyses show that costs of resistance are not necessary to make adaptive therapy superior to higher dose treatment~\cite{strobl2020turnover, viossat2021theoretical}. Nevertheless, such costs -- which could be exacerbated by auxiliary treatments~\cite{silva2010theoretical, silva2012evolutionary} -- are typically predicted to amplify clinical gains~\cite{viossat2021theoretical}.

Lastly, is competition the only important ecological interaction? Studies {\em in vitro} and in mice have detected positive ecological interactions (mutualism and commensalism; reviewed in~\cite{Tabassum2015}) and asymmetric interactions (parasitism) between cancer clones~\cite{Miller1988, Noble2021_Paracrine}. These observations suggest that our theoretical models of clonal dynamics during cancer treatment may be overly simplistic, and they underscore the need for more and better data.

\subsection{What is the role of plasticity and drug-induced mutations in adaptive therapy?}\label{section_plasticity}
\color{red}Eunjung Kim: \color{black}
The effectiveness of treatment holidays drastically changes when considering phenotype switching between drug-sensitive and -resistant phenotypes. Plasticity is often modeled as the expression of resistant cellular traits which vary from completely sensitive to fully resistant~\cite{clairambault2019survey, clairambault2019evolutionary}, in multi-dimensional fashion to consider multi-drug resistance~\cite{cho2018modeling, cho2018Bmodeling}. Treatment breaks can halt the expansion of the resistant cell population facilitated by drug induced mutations or phenotype switching from sensitive to resistant states during therapy. Since phenotype switching to resistant states is often reversible (reviewed in~\cite{Boumahdireview}), treatment holidays have the potential to re-sensitize the resistant cell population to future drug rechallenges. A recent experimental study demonstrated that gene expression in melanoma cells reversed during treatment holidays, causing the cells tp re-sensitize to a BRAF inhibitor rechallenge~\cite{Kavran}.

The switching rate from resistant to sensitive states can impact the benefit of adaptive therapy. Tumor dynamics described by both competition and phenotypic plasticity predict that adaptive therapy (dose skipping) outperforms standard of care at different degrees among patients with advanced metastatic melanoma~\cite{kim2021adaptive, masud2022containing}. Among parameters that govern treatment response dynamics, both the switching rate from resistant to sensitive states and the growth rate of sensitive cells determines the benefits of adaptive therapy. In another study, a fixed schedule intermittent therapy was predicted to outperform the standard of care when treatment could induce resistant mutations in the cells~\cite{greene2019mathematical}. These properties of tumor plasticity or drug-induced mutation are variable between cancer types and possibly vary between cancer cells. For example, in melanoma, it was shown that phenotypic plasticity is more evident in one cell line than another~\cite{smalley2019leveraging}. There may be even more variability across patients in terms of how resistance emerges and is maintained. Thus, identifying the presence of phenotypic plasticity in a specific tumor could be an important factor in deciding if and how adaptive therapy should be applied.

\begin{figure}[t!]
    \centering
    \includegraphics[width=1\linewidth]{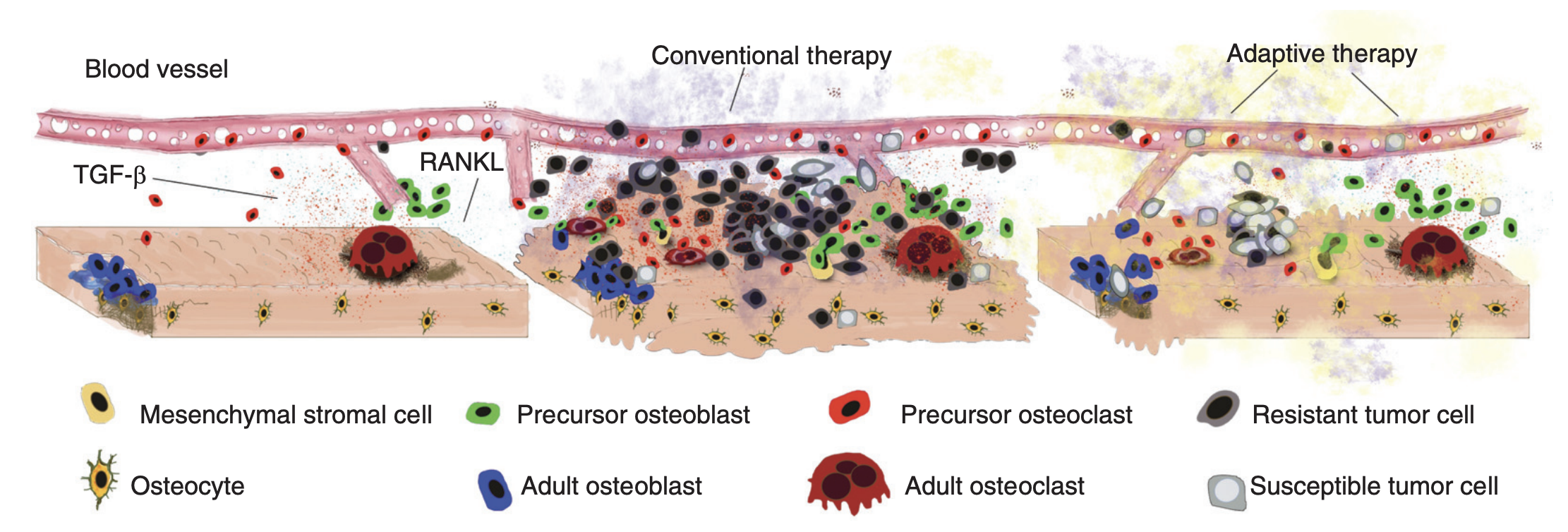}
    \caption{{\bf Disruption and restoration of tissue homeostasis}:  Figure reproduced from ref.~\cite{basanta2017homeostasis}. Left: bone tissue homeostasis, including bone resorption by osteoclasts and osteoblasts. Middle: Tumor cells cause disruption of homeostasis, leading to altered microenvironment factors. Conventional therapy leads to increasing tumor resistance. Right: Evolution-based treatment strategies aim to restore some degree of homeostasis while allowing the tumor to remain sensitive to future treatment.}
    \label{fig:homeostasis}
\end{figure}
    
\subsection{What is the role of homeostasis and normal tissue in adaptive therapy?}
\color{red}David Basanta: \color{black} A feature of current models of adaptive therapy lies in their simplicity in terms of algorithms and assumptions. One key simplification is that tumor heterogeneity can be reduced to  the types of cancer cells such as sensitive cells (that pay a significant fitness cost during treatment) and treatment-resistant cells (that may incur a cost of resistance relative to sensitive cells). In reality, the fitness of a cancer cell is not simply a cell-intrinsic property but includes its ability to take advantage of its dynamic tumor environment that includes not just other cancer cells but normal cells, vasculature, immune cells and extra-cellular matrices (see figure \ref{fig:homeostasis}).

Far from living in isolation, cancer cells colonize tissues with an existing ecosystem made up of healthy, stromal cells that communicate with each other via molecular factors in order to maintain homeostasis. The tissue  has a distinct physical and spatial architecture~\cite{basanta2013exploiting}. In the bone for instance, homeostasis results from the interactions between several cell types such as osteoblasts, osteoclasts, osteocytes, monocytes, macrophages and mesenchymal stem cells (MSC)~\cite{bussard2008bone}. While metastasis is a highly inefficient process, successful colonization of the bone by metastasizing prostate cancer cells leads to a process called the vicious cycle~\cite{esposito2018biology, cook2014integrating}. Successful prostate cancer cells in the bone take advantage of the interactions and signaling that goes on between the normal cells as they maintain homeostatic tissue microenvironments~\cite{basanta2013exploiting}. Factors released by normal cells such as Transforming Growth Factor $\beta$ are utilized by nearby prostate cancer cells. Such factors accelerate the proliferation and survival of the cancer cells.

Proximity to stromal cells provides other benefits to cancer cells undergoing treatment. For instance, bone metastatic prostate cancer cells near MSCs are pre-selected to possess some level of chemoresistance~\cite{mcguire2021mesenchymal}. Such cells are primed for resistance even prior to treatment exposure. Also in the bone, myeloma cells (bone cancer) near MSCs or in the presence of growth factors secreted during bone remodeling can survive standard of care treatments based on proteosome inhibitors like bortezomib~\cite{xu2012bone}. This environmentally mediated drug resistance (EMDR) explains why anti-cancer treatments prove less effective than otherwise expected. In the context of adaptive therapy, EMDR may provide the cancer cells with a therapy refuge regardless of whether they are resistant or sensitive to treatment. 

For cancers where EMDR plays a large role, it may be advisable to be more aggressive in applying therapy and reducing the tumor burden.  This is because treatment sensitive cells will remain viable even after large reductions in tumor burden.  These sensitive cells surviving in or near a stromal refuge can then provide a source of competition for resistant cancer cells during drug holidays.  Furthermore, one can include adjuvant treatments, targeting stromal cells, to modulate the role of EMDR.  The goal of EMDR modulating drugs would not be to maximize tumor kill.  Rather, such modulation would aim to improve the efficacy of adaptive therapy approaches to better control tumor burden and maintain quality of life over long periods of time.

Future adaptive therapy models should allow for EMDR.  Such models could more faithfully incorporate the tumor microenvironment and the role of normal cells. Such models could then evaluate how best to manage or exploit EMDR when designing adaptive therapy protocols~\cite{masud2021impact}. Even more might be gained by developing models that also recapitulate tissue homeostasis prior to carcinogenesis. As discussed elsewhere~\cite{basanta2017homeostasis}, cancers initiate in normal tissue environments and progressively overcome and exploit the rules of homeostasis. Adaptive therapy aims to control the tumor by introducing a different type of homeostasis. Hence, in improving the original adaptive therapy algorithm, we should consider the homeostasis that was disrupted by the tumor as well as the homeostasis that might be engineered by therapy.

\section{Design and validation of dosing protocols}
\subsection{Cure or control?}\label{control_vs_cure}
\color{red}Jill Gallaher: \color{black}
An explicit  goal of adaptive therapy is to turn cancer into a chronic disease with sporadic (but life-long) management. Taking this approach likely means abandoning the hope of cure. Thus, its current appeal and modeling contexts have been for patients with essentially no curative options. However, there are cases when adaptive therapy would have been preferred when standard of care results in recurrence and adaptive therapy would have either resulted in longer control or less dose and therefore better quality of life. In other cases, standard of care could result in cure or longer control. But is identifying such patients prior to treatment even possible?  Furthermore, if the standard of care regimen is tried and cure does not result, then it may not be possible to switch to an adaptive therapy regimen because at that point the resistant population of cancer cells may be too large compared to the sensitive cells to establish sufficient control~\cite{mcclatchy2020modeling}. The window for extended disease control using an adaptive therapy protocol may only be open prior to treatment. The decision must be made at the start. If one only opts for standard of care for cure, failed curative attempts could lead to reduced survival times~\cite{Monro2009} relative to adaptive therapy. So how does one decide between these opposing strategies? What key disease characteristics are needed to stratify patients into a treat-to-cure cohort (using continuous therapies) versus a treat-to-contain cohort (using adaptive therapies)~\cite{hansen2020cancer}?

Towards deciding between treat-to-cure or treat-to-contain strategies, bacterial studies in addition to cancer studies provide some insights. Initial pathogen heterogeneity may favor adaptive therapies. The treatment response of heterogeneous bacterial colonies depends on competition and tradeoffs among the bacterial strains. Colijn et.al. studied how competition between microbial populations affects the optimal dosing of antibiotics~\cite{colijn2015competition}. Large antibiotic doses were in some cases observed to reduce the bacterial load and prevent resistance and in other cases to select for more antibiotic resistant cells. They found that even if aggressive treatment was optimal for individual strains, moderate treatments were better to avoid resistance for the entire community when there was strong inter-strain competition~\cite{colijn2015competition}. Hansen et. al. proposed a balance threshold hypothesizing that containment delays progression only if the overall effect of competitive suppression exceeds the overall effect of mutational input~\cite{hansen2020cancer, hansen2017use}. In cancer, like bacteria, tradeoffs between therapy resistance and competitiveness in the absence of therapy favor adaptive therapies over treat-to-cure. But these tradeoffs among cancer cells may only manifest in specific contexts or through temporary constraints on cell functions~\cite{strobl2020turnover}.  A retrospective analysis of the Zhang et. al. metastatic prostate cancer trial found that the success of long-term survivors of adaptive therapy resulted from a large and asymmetrical competitive effect of sensitive cells on resistant cells~\cite{zhang2022evolution}. Such will not always be the case. The effect of competition and tradeoffs may lessen if cells are not in direct contact~\cite{strobl2021spatial}. For example, if the tumor is very invasive or if different metastases have different compositions of sensitive and resistant cells, the effects of tradeoffs and competition are reduced~\cite{gallaher2018spatial, gallaher2022sum}. Diverse metastatic sites or spatial segregation of heterogeneous cancer cells within tumors mean that different tumors or locations within tumors may respond strongly to treatment while others grow unimpeded. In these cases, an attempt at cure might be better than containment~\cite{gallaher2022sum}.

The ratio of sensitive to resistant subpopulations as well as the transition rates between them influence the efficacy of adaptive therapy and continuous treat-to-cure therapies. Both benefit from there being a low frequency of resistant cancer cells. But, adaptive therapy may benefit more. An aggressive, maximum-tolerated chemotherapy approach will result in a larger initial response to therapy~\cite{hansen2020modifying}, but if there is not complete disease eradication, even a small or emergent population of resistant cancer cells can guarantee eventual disease progression. High doses necessarily promote the competitive release of pre-existing resistant cells. However, high doses may also prevent de novo drug resistance by lessening the pool of sensitive cells from which resistant cells emerge~\cite{colijn2015competition}. Phenotypic plasticity poses another challenge for both therapeutic strategies where the application of treatment accelerates the transition of cancer cells into resistant states (see section \ref{section_plasticity}). Nevertheless, adaptive therapy may be favored when treatment breaks result in the re-sensitization of cancer cells to the drug.  In that case, Kim et. al. found that in melanoma patients a high rate of re-sensitization is required for either treatment option to work well, but that the optimal treatment strategy, as expected, depends on competition, tradeoffs and the initial fraction of resistant cancer cells~\cite{kim2021adaptive}.

Overall disease burden and initial response to therapy should influence the choice of cure versus containment. In the case of anti-microbial drugs, Kouyos et. al. evaluate the decision for aggressive versus moderate dosing in terms of two opposing ecological and evolutionary processes~\cite{kouyos2014path}.  Ecologically, the rate of disease burden reduction will increase with increasing dose.  Evolutionarily, increasing the dose increases the selection pressure for resistant microbes~\cite{kouyos2014path}. Their work concludes that the optimal dose should be high enough to reduce the patient’s disease burden, and low enough to forestall the emergence of resistance.  Adaptive therapy attempts to break this constraint by applying high doses when the frequency of sensitive cells is high and removing therapy when this frequency has declined. Yet, successful adaptive therapies may mean maintaining large tumor burdens. A proposed modification of adaptive therapy suggests that increasing the baseline tumor burden of smaller tumors (if it is safe to do so) would help increase competition effects~\cite{hansen2020modifying}. While letting the tumor burden become higher prior to reinitiating therapy prolongs the period during which sensitive cells outcompete resistant ones, it can bring unintended negative consequences. The patient must be able to tolerate the large burden without debilitating or life-threatening symptoms. Also, maintaining a high tumor burden may increase the likelihood of new metastases or evolutionary breakthroughs by the cancer cells. 

The tumor burden alone does not reveal the underlying dynamics of cell turnover~\cite{gallaher2019impact}, which is also important for treatment response. This background turnover results from the replacement through proliferation of cells that regularly die from spontaneous apoptosis, an immune response, or lack of resources. Both therapeutic strategies can benefit from a higher background cell turnover rate.  An aggressive treatment may substantially decrease the tumor burden and increase the probability of cure by increasing the chance of spontaneous death of the resistant population~\cite{strobl2021spatial}.  An adaptive therapy approach benefits from a high cell-turnover rate by increasing the competition to mutational input balance threshold~\cite{hansen2020cancer}.  For adaptive therapy, a high cell turnover rate during the period when therapy is off increases the opportunities for sensitive cells to replace resistant cancer cells. 

There are also practical clinical considerations of using each treatment strategy. An adaptive protocol must have frequent measures on which to base the decisions of when to increase or decrease dose rates. For adaptive therapy in prostate cancer, PSA is used as a surrogate for burden. In other cancers it may be imaging, ctDNA, or other molecular markers. The biomarker used for decision-making needs to accurately measure changes in the disease burden and state. Frequent measures are best, thus inexpensive and less invasive biomarkers are favored. Ideally, decisions for adaptive protocols could also be guided by measurements of drug resistance, evolvability, or competition if possible. Otherwise, these measurements might be used as stratification factors from pre-treatment tissue biopsy. A short induction period to determine disease kinetics could help with the stratification of patients with higher or lower likelihoods of cure under an aggressive treatment strategy. This allows for some measures of the cancer’s eco-evolutionary dynamics without committing to a specific therapeutic regimen~\cite{mcclatchy2020modeling}. Further, overall survival is an important measure for comparing treatment strategies, but it must be balanced with toxicity and quality of life~\cite{milano2020oligometastases}. For successful adaptive therapy, drug timing, which includes pharmacokinetics and pharmacodynamics, must be aligned with the growth rate of the tumor and the accumulating side effects for the patient. Attempting to cure a tumor with a slow response means a longer application of aggressive treatment, so drug toxicity becomes a key consideration. With adaptive therapy the treatment breaks can improve quality of life and reduce overall dose rates, but there is potential for accumulating doses and side effects from subsequent treatment cycles over an indefinite course of therapy. Cure or control could be favored depdending on the patient, the disease state, and the drugs used.

\subsection{What is the optimal adaptive dose administration protocol?}
\color{red}Yannick Viossat: \color{black}
Adaptive therapy often refers to the specific protocol used in the initial prostate clinical trial~\cite{zhang2017integrating}. However, the concept has wider applicability~\cite{gatenby2009adaptive, enriquez2016exploiting, bacevic2017spatial, Carrere2017, gallaher2018spatial, viossat2021theoretical, cunningham2018optimal, cunningham2020optimal, hansen2020modifying}. The prostate trial's design was driven by a compromise between mathematical model results and clinically feasible treatment protocols. In this section, we review optimal protocols revealed through investigations of mathematical models. Subsequent sections review the best practices to incorporate experimental (section \ref{maxi_section}) and clinical data (section \ref{renee_section}) relating to dose modulation protocols.

Many models emphasize competition between sensitive and highly resistant cells and assume that the larger the tumor size, the stronger the competition~\cite{Martin1992a, Monro2009, zhang2017integrating, Carrere2017, carrere2020stability, martin1992b, strobl2020turnover, viossat2021theoretical}. Such models suggest maintaining the tumor at the maximal acceptable size, in order to maximize competitive suppression of resistance~\cite{hansen2020modifying, viossat2021theoretical}. This may require delaying treatment if the tumor is initially small~\cite{Monro2009, cunningham2020optimal, hansen2020modifying, viossat2021theoretical}. Time to tumor progression may be delayed by switching to high doses a short time before containment fails~\cite{wang2021optimizing, viossat2021theoretical}, but at the risk of making the tumor less treatable afterwards~\cite{viossat2021theoretical}. Moreover, aggressive treatment may increase toxicity and drug-induced mutations~\cite{kuosmanen2021drug}. However, maintaining a substantial tumor burden may create other potential problems: a lower quality of life, more metastases, increased mutation from sensitive to resistant tumor cells~\cite{Martin1992a, hansen2017use}, or the appearance of new tumor cell types. Some models with a death-rate that increases proportionally with tumor size find that MTD is superior to adaptive therapy (dose skipping)~\cite{mistry2020evolutionary}. 

If resistant cells are only partially resistant, they may be targeted by treating mildly, to exploit competition with sensitive cells, or aggressively, to exploit their remaining sensitivity. A sensible strategy is to first exploit competition by stabilizing tumor size, but then switch to MTD long before stabilization fails, as opposed to stabilizing the tumor for as long as possible when resistant cells are fully resistant~\cite{hansen2020modifying, viossat2021theoretical}. The switching time could be timed with a decrease of treatment efficiency~\cite{wang2021optimizing}, but the practical implementation and whether similar conclusions hold for models with many types of tumor cells~\cite{Pouchol2018} remains to be investigated. 

Tumor stabilization may be achieved through dose-modulation or dose-skipping~\cite{enriquez2016exploiting, gallaher2018spatial}. If competition increases with tumor size, a continuous low dose treatment maintaining the tumor at a given threshold may be preferable to an intermittent treatment maintaining it between this threshold and some lower size. This is not a compelling argument against intermittent treatments though: an intermittent treatment maintaining tumor size between this threshold and some \emph{larger} size may increase competition even more~\cite{hansen2020modifying, viossat2021theoretical}. Tumor stabilization might also normalize tumor vasculature, leading to a larger drug efficiency~\cite{gatenby2009adaptive, enriquez2016exploiting}, and dose-modulation might normalize tumor environment more than dose-skipping~\cite{enriquez2016exploiting}. Evidence remains scarce. In a mouse model~\cite{enriquez2016exploiting}, a dose-modulation strategy was more effective than a dose-skipping strategy, but maybe because the cumulative dose in the dose-modulation arm was higher. 

A tumor may be temporarily stabilized by a constant dose treatment during a short time interval. A practical question is to determine the appropriate stabilization dose. Cunningham et. al.~\cite{cunningham2020optimal} found that an upward dose-titration protocol, gradually increasing the dose until the tumor is stabilized, works better than a dose reduction protocol. Indeed, starting from a high dose may quickly select for resistant phenotypes. However, with upward titration, tumor size may become dangerously large before the dose is sufficiently high to stabilize it. Rather than a fixed dose modulation (e.g. 10\%), weighting the dose according to tumor responsiveness may lead to quicker stabilization~\cite{viossat2021theoretical}. Moreover, some protocols keep the same dose if tumor size changes little since the previous measurement~\cite{gatenby2009adaptive, enriquez2016exploiting, gallaher2018spatial}. Successive small changes, with a large total effect, may then never trigger dose-modulation. This suggests the next dose should depend on the most recent change in tumor size but perhaps also on its absolute size compared to some target~\cite{viossat2021theoretical}. Many of these ideas remain to be empirically tested and may be difficult to implement clinically.

Finally, agent-based models~\cite{you2017spatial, bacevic2017spatial, gallaher2018spatial, strobl2021spatial} allow for testing of features that are not easily incorporated into differential equation models: spatial structure, cell mobility, or quiescence. Spatial structure may increase the cost of resistance, as resistant cells may be trapped inside the tumor, far from the proliferative edge. These models also lead to observations that are not easy to understand theoretically, such as the greater efficiency of dose-skipping over dose-modulation in \cite{gallaher2018spatial}. This highlights that simple models may miss important phenomena and that more data and modeling are needed to optimize adaptive therapies.

\subsection{How can we leverage mathematical modelling to support testing of adaptive therapy in the wet lab?} \label{maxi_section}
\color{red}Maximilian Strobl: \color{black} Thanks to promising pre-clinical and clinical results, there is growing interest in extending adaptive therapy to new disease settings. To do so requires experimental platforms for testing and, if necessary, improving the safety and efficacy of adaptive protocols. Experimental systems are models and come with inherent assumptions and limitations. Mathematical modelers and experimentalists should collaborate closely in order to design pre-clinical studies to validate theoretical models, assert safety, and develop adaptive protocols with the maximum benefit to patients.

Surveying the experimental literature on adaptive therapy, and based on our own experience, we identify three areas in need of further research. First, how do we design experiments to assess the competitive suppression in a particular cancer and thus the scope for such patients to benefit from adaptive therapy? To-date, experiments have employed one of three model systems, or combinations thereof: i) 2-D \textit{in vitro} cell culture (e.g. refs.~\cite{silva2012evolutionary, bacevic2017spatial, farrokhian2022measuring, nam2020dynamic, bondarenko2021metronomic}), ii) 3-D \textit{in vitro} spheroids (e.g. refs.~\cite{bacevic2017spatial, strobl2020turnover, bondarenko2021metronomic}), and iii) orthotopic {\em in vivo} mouse models, in which human cells are injected into immuno-compromised animals (e.g. refs.~\cite{gatenby2009adaptive, enriquez2016exploiting, smalley2019leveraging, wang2021optimizing, wang2021fixed}). 2-D and 3-D \textit{in vitro} models are inexpensive and quick, and allow for easy manipulation and monitoring of the ``tumor.'' In contrast, by incorporating vasculature and stroma, orthotopic mouse models are more realistic, but they are expensive. Mouse models often do not include an immune system and see human cells competing with mouse rather than human cells. A solution to this problem will be the use of more advanced technology, such as organ-on-chip models~\cite{kashaninejad2016organ, wang2021considerations} or spontaneous mouse models, where mouse tumors develop ``naturally'' in their tissues of origin~\cite{kersten2017genetically, cespedes2006mouse}. But even with more advanced experimental systems, limitations remain. To address these, we need to better understand what cells compete for (as discussed in Section 1.2.), and how we can best quantify this competition (e.g. the ``game assay''~\cite{kaznatcheev2019fibroblasts,farrokhian2022measuring}). We propose that by playing out different scenarios \textit{in silico}, mathematical models can help us to refine what experiments we should perform, and in what experimental system(s), in order to deduce the competitive landscape in tumors and in order to inform on how adaptive therapy will perform in patients.

Second, there is the question of how drug-resistance is modeled in the wet lab. One approach evolves resistant cells through long term drug exposure, and subsequently performs experiments by mixing these cells with parental, sensitive cells (e.g. refs.~\cite{bacevic2017spatial, nam2020dynamic, wang2021optimizing}). This has the advantage that the resistant population can be characterized (e.g. measuring its growth rate), the initial resistant cell fraction can be controlled, and cell populations can be fluorescently tagged and followed over the course of the experiment. However, this design implicitly assumes the pre-existence of resistant cells in the tumor and neglects the role of plasticity. Alternatively, drug resistance can be allowed to evolve naturally over the course of the experiment (e.g. refs.~\cite{gatenby2009adaptive, enriquez2016exploiting, smalley2019leveraging}). But such experiments are time consuming and provide less information about the resistant population. We suggest using mathematical modelling to examine how best to leverage each of these two approaches to gain information on the cancer of interest. In addition, emerging clone-tracking technology allows for ever more in-depth study of tumor evolution~\cite{morgan2021applications}. Mathematical modeling will be a useful tool in designing, and interpreting results from, clone-tracking experiments~\cite{acar2020exploiting, damaghi2020harsh, johnson2020integrating}.

Finally, there is the question of how to translate treatment algorithms from mathematical or experimental models into clinical practice. Most mathematical models of adaptive therapy neglect drug pharmacokinetics, but clearly this impacts the drug delivery to the tumor and differs between animals and patients. In addition, there is a question of time scales: how does a weekly follow-up in mice compare to a re-assessment every 3 months in patients? And, what happens when treatment cannot be adjusted as planned due to toxicity or practical constraints (e.g. machine failure, or the intended day falling on a holiday/weekend)? This raises the question of how robust are adaptive schedules to deviations, and what is the best strategy with which to respond when deviations occur. Some initial work on this topic has been carried out~\cite{dua2021adaptive, wang2021fixed} and we encourage more research in this direction in order to inform experimental and clinical trial design.

\subsection{What are best practices to design adaptive algorithms for multiple drugs?}
\color{red}Jeffrey West: \color{black} It remains unclear how to extend adaptive therapy approaches to multiple treatments. When multiple drugs are available, the combinatorial possibilities expand rapidly. With $n$ treatments available, there exist $2^n$ possible combinations, each of which may be administered at each treatment decision point. Current adaptive trials often utilize less than the full range of $2^n$ combinations. For example, the metastatic castrate-resistant prostate cancer adaptive trial (NCT02415621) administers Lupron as a continuous backbone while abiraterone is given adaptively. The advanced BRAF-mutant melanoma adaptive trial (NCT03543969) administers encorafinib and binimetinib in combination adaptively, with nivolumab administered continuously. In both examples, opening the trial design to include the full range of treatment permutations may extend therapeutic control, but at the cost of computational and investigational complexity.

Recently, the concept of steering tumor dynamics into periodic, repeatable evolutionary cycles was proposed~\cite{newton2019nonlinear, ma2021role, liu2022identifying}. The ordering and timing of treatment combinations is chosen carefully to drive tumor phenotypic composition into a ``cycle'' such that tumor composition at the start and end of a cycle of therapy are  approximately equivalent~\cite{west2020towards, dua2021adaptive}. Evolutionary cycling was implemented as a strategy to combat resistance to osimertinib (a third-generation tyrosine kinase inhibitor) in EGFR-mutant non-small cell lung cancer~\cite{wang2021fixed}. Dynamics were described by a Lotka-Volterra competition model within a nonlinear mixed-effects modeling framework, and potential treatment schedules were screened in silico to select fixed protocols that drive tumor dynamics into periodic cycles. These fixed treatment plans implemented {\em in vivo} outperformed standard of care treatment schedules in a majority of cases. This study and others~\cite{wang2021optimizing, thomas2022silico, west2020towards, west2019multidrug} illustrate the feasibility of model-driven treatment planning to reduce the combinatorial complexity for multi-drug adaptive therapies.

An alternative multidrug adaptive approach is to identify collaterally sensitive treatments such that resistance to first line treatment induces sensitivity to secondary treatments~\cite{silva2010theoretical, basanta2012exploiting}. In double-bind therapy, the drugs are given sequentially rather than in combination~\cite{gatenby2009lessons}. Using mathematical models, it is possible to determine the optimal switching time between a pair of collaterally sensitive drugs~\cite{yoon2018optimal}. This approach can be extended to infer the optimal timing and ordering of a set of collaterally sensitive drugs~\cite{yoon2021theoretical}, which can allow for evolutionary steering~\cite{iram2021controlling, acar2020exploiting} or even extinction~\cite{gatenby2020eradicating, gatenby2019first}. A similar strategy termed the ``primary-secondary'' approach, administers the primary treatment adaptively, while the secondary treatment is administered within each adaptive treatment cycle in order to suppress long-term resistance~\cite{west2019multidrug}.

Another alternative multi-drug approach known as extinction therapy may provide a way out of the control versus cure conundrum introduced in section \ref{control_vs_cure}.  For many incurable cancers or specific patients that failed to be cured, an aggressive therapy given continuously will generate a complete response rendering the cancer temporarily clinically undetectable sometimes for periods of years, other times for just months.  Rather than wait for the period of remission to end before switching therapies, extinction therapy aims to exploit the vulnerabilities of small, fragmented populations~\cite{artzy2021novel, johnson2019cancer, konstorum2016feedback}.  In this case these small populations are the remnants that survived therapy either by virtue of resistance or position within the tumor (sometime referred to as stromal protection when the structure of normal cells prevent therapy reaching cancer cells).  In models of extinction therapy, the initial therapy (called the first-strike) is stopped as soon as the disease burden shows a complete response~\cite{gatenby2019first, gatenby2020integrating}.  At this point, therapy becomes a sequence (e.g. 45 – 90 days per sequence) of second strikes using different drugs with different modes of action, and that will not generate undue toxicities.  While untried, one can imagine starting a patient that might be cured with an aggressive therapy.  If this therapy only generates a partial response then immediately switch to another drug and/or an adaptive therapy before disease progression.  If the therapy produces a complete response then go into an extinction therapy regimen aiming for cure.  If permanent remission does not ensue then the initial first strike drug likely is still effective, and can then be used for an adaptive therapy.  By switching therapies sooner before complete resistance has evolved the physician and patient retain the option for switching to an adaptive therapy.  While models of extinction therapy have been developed~\cite{gatenby2020eradicating}, clinical evidence is sparse but supported by the standard of care multi-step curative treatment in Pediatric Acute Lymphocytic Leukemia~\cite{li2022antibody}, by two case studies involving cure in patients with metastatic breast cancer~\cite{chue2019can, chue2019case}, and an ongoing clinical trial for patients with pediatric rhabdomyosarcoma~\cite{reed2020evolutionary}.
    
\section{Challenges and opportunities in clinical translation}
\subsection{Is real-time patient prediction feasible?}\label{renee_section}
\color{red}Renee Brady-Nicholls: \color{black} 
Predicting precisely when a patient will progress during adaptive therapy offers the opportunity to appropriately modulate treatment, thereby extending patient response and survival. This requires sufficient monitoring of an individual patient’s disease using appropriate clinical markers. Choosing an appropriate biomarker depends on the extent of the disease (e.g. localized versus metastatic, or hormone sensitive versus castration resistant prostate cancer), as well as how frequently said biomarker can be collected to adequately follow the disease trajectory.

Prior to making model predictions, the chosen model should be calibrated and validated to demonstrate that it can accurately describe patient-specific biomarker dynamics~\cite{brady2019mathematical}. This requires analyzing the model to determine the sensitivity and identifiability of model parameters, relative to the data. Given a chosen model and available data, model parameters may be difficult to accurately estimate. This might be due to the model complexity or structure, as well as the given data. Evaluating the sensitivity of the model outputs, in this case the change in a modeled biomarker over time, with respect to small perturbations in the model parameters identifies sensitive and insensitive parameters~\cite{banks2009}. Sensitive parameters should be evaluated for correlations with other parameters, as correlated parameters should not be estimated concurrently. If two parameters can be uniquely identified, then they are said to be identifiable~\cite{olufsen2013} and parameter optimization techniques can be used to determine their optimal values relative to the given data. Appropriate model development, calibration, and validation are essential when developing predictive models. If the model cannot accurately describe the data, then it should not be used to make predictions of subsequent patient responses.

\begin{figure}[t!]
    \centering
    \includegraphics[width=0.6\linewidth]{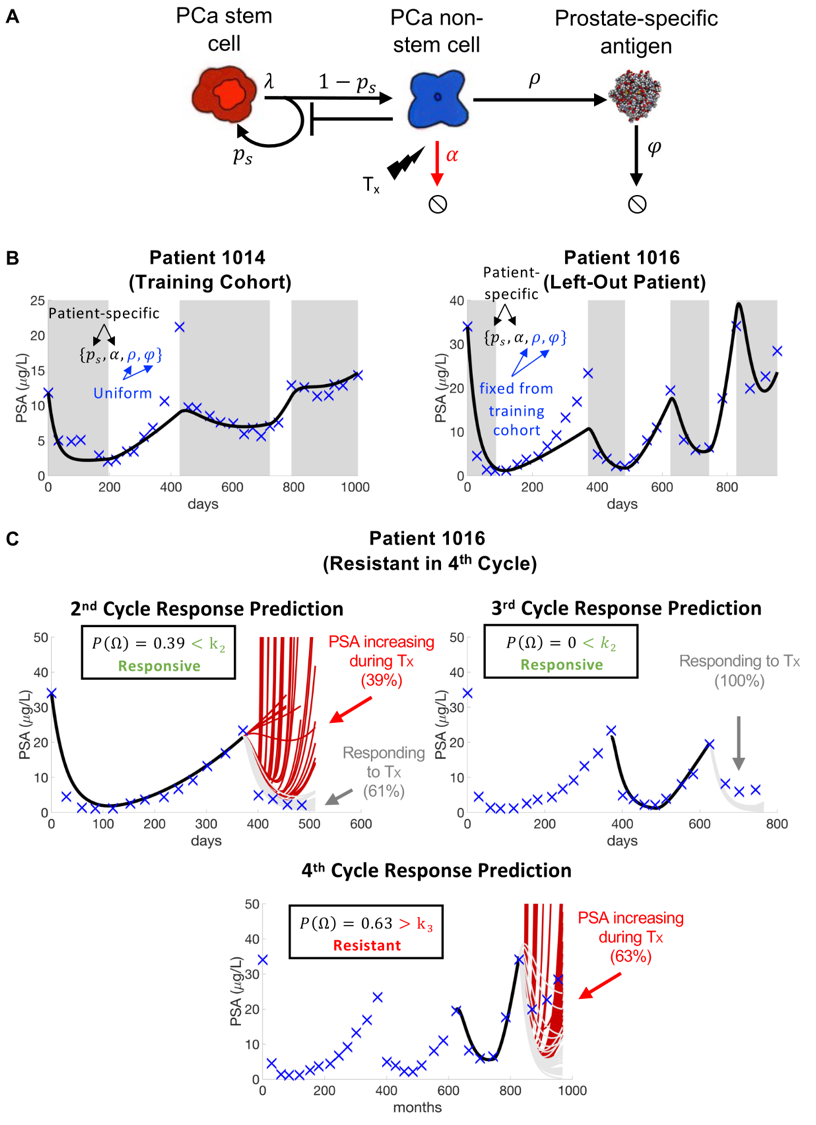}
    \caption{{\bf Model schematic, calibration, validation, and prediction}: A) Model schematic of treatment-resistant stem cells, sensitive non-stem cells, and prostate-specific antigen interactions. (B) Model calibration (Patient 1014) and validation (Patient 1016). Nested optimization was used to determine the cohort uniform parameters $\rho$ and $\varphi$ and the patient-specific parameters $p_s$ and $\alpha$ for the training cohort. The uniform values were fixed in the testing cohort and optimization was used to find the patient-specific parameters $p_s$ and $\alpha$. (C) Model predictions for Patient 1016. The model predicted resistance in 39\% of cycle 2 simulations and response in 100\% of cycle 3 simulations. Cycle 4 predictions showed resistance in 63\% of model simulations.  Using cycle-specific cutoffs $k_2,k_3$ and $k_4$, the model correctly predicted that Patient 1016 would continue to respond in cycles 2 and 3 but become resistant in cycle 4.}
    \label{fig:prediction}
\end{figure}

Figure \ref{fig:prediction} illustrates a case-study of the feasibility of real-time patient prediction. Here, the longitudinal biomarker known as prostate-specific antigen (PSA) is used to fit a mathematical model and provide a patient-specific prediction in response to adaptive therapy in metastatic castration resistant prostate cancer~\cite{brady2021predicting}. Despite much controversy surrounding the clinical use of the absolute value of PSA in both the detection and monitoring of prostate cancer~\cite{Kim2015}, it has been shown to be an effective and inexpensive way to follow a patient's response trajectory over time. Many mathematical models based on a variety of plausible mechanisms have been developed to describe PSA dynamics in response to treatment~\cite{zhang2017integrating, Baez2016, hirata2010development, morken2014mechanisms, portz2012clinical}. 

The mathematical model (Figure \ref{fig:prediction}A) uses stem and non-stem cell dynamics to describe patient-specific PSA dynamics in response to treatment. The model has five parameters ($p_s$, $\lambda$, $\alpha$, $\rho$, $\varphi$). Sensitivity analysis found that $\lambda$ was insensitive, while identifiability analysis showed that $p_s$, $\alpha$, $\rho$, and $\varphi$ were uncorrelated and identifiable. A leave-one-out analysis was used to calibrate and validate the model to longitudinal PSA data from 16 patients. That is, nested optimization was used to estimate patient-specific parameters for $p_s$ and $\alpha$, and uniform parameters for $\varphi$ and $\rho$ to accurately describe individual patient data for the 15 patients in the training cohort. The uniform $\varphi$ and $\rho$ values were then fixed and optimization was used to find the $p_s$ and $\alpha$ values for the left-out patient (Figure \ref{fig:prediction}B). Once calibrated and validated to the patient data, the model was used to make patient-specific response predictions. Parameter analysis identified the stem cell self-renewal rate $p_s$ as the primary driver of differences in treatment response dynamics between responsive and resistant patients. This parameter was used to make subsequent response predictions. That is, the distribution of changes in $p_s$ from treatment cycle $i$ to $i+1$ was used to predict an individual patient's response in cycle $i+1$ (Figure \ref{fig:prediction}C). The model was able to predict patient response with 81\% accuracy~\cite{brady2021predicting}. A similar modeling approach was used in biochemically recurrent prostate cancer patients receiving intermittent androgen deprivation therapy to predict response dynamics with 89\% accuracy~\cite{brady2020prostate}.

We can learn several lessons from these studies when applying real-time prediction of adaptive therapy in new diseases. Model-predictions are dependent on the quality and time-resolution of patient-specific biomarkers. Alternative biomarkers such as circulating tumor DNA (ctDNA)~\cite{hennigan2019, ku2019, lau2020}, circulating tumor cells (CTCs)~\cite{ried2020, salami2019}, and relatively new biomarkers such as urine~\cite{lemos2019, tosoian2021} have been shown to be prognostic in prostate cancer and other diseases. Like PSA, these markers can be collected relatively frequently and via minimally-invasive methods. They can be used to develop appropriate models that can be used to predict response to adaptive therapy.


\subsection{Do adverse effects of maintaining high tumor burden negate potential benefit?}
\color{red}Joel Brown: \color{black} 
The prostate adaptive trial (NCT024515621) is instructive here.  The original model imagined two categories of sensitive cells that both require testosterone~\cite{zhang2017integrating}. One producing its own, and the other requiring exogenous testosterone.  Resistant cells are independent of testosterone and hence unaffected by androgen-focused therapy.  In this model the cost of resistance was assumed to occur primarily through carrying capacity with some contribution of competition coefficients.  

The adaptive therapy patients performed better than the contemporaneous controls (median of 33.5 and 14.3 months radiographic progression free survival, respectively)~\cite{zhang2021response}. To be considered for the trial, an initial response to therapy of at least a 50\% decline in PSA and putative tumor burden was required.  This meant they were responders and hence enjoyed a better prognosis than non-responders, and non-responders were not eligible for this trial~\cite{mistry2021reporting}.  The contemporaneous control group was selected from patients receiving continuous therapy and who, like the trial patients, were responders.  No non-responders were included in the control cohort.  By way of caveat, this is an important consideration for any trial of adaptive therapy that lacks double-blind, randomized control arms.  Most strikingly, some of the adaptive therapy patients performed better than expected from the initial model.  At time of writing, 4 men remain on adaptive therapy after 4 – 6 years.

Theory tells us that adaptive therapy, or more specifically competitive release therapy, works best if the sensitive cells are allowed to grow to the point where both frequency- and density-dependent feedbacks are strong.  The frequency-dependent effects offer hope that the sensitive cells are competitively superior to the resistant ones.  This cost of resistance can manifest as a reduction in maximum growth rates, reduction in carrying capacity or maximal cell density, and the competitive effect of each cell type on the other~\cite{zhang2022evolution,strobl2020turnover,pressley2021evolutionary}. A retrospective analysis of the near-finished trial~\cite{zhang2022evolution}, permits parameter estimations, some patient specific, others estimated as patient-wide parameters. The efficacy of adaptive therapy seems best explained by highly asymmetric competition.  It was estimated that the sensitive cells (a streamlined model combining the two sensitive cell types into one) have a competitive effect (per cell) on resistant cells that is six times higher than the reverse effect of resistant cells on sensitive ones.  As others have noted in their models, such a high asymmetry favoring sensitive cells relative to resistant cells can result in indefinite control and disease containment~\cite{viossat2021theoretical}.

The prostate adaptive trial does provide some evidence that prolonged sensitivity may outweigh adverse effects of maintaining high tumor burden, but the promise of being able to perpetually control each patient’s disease did not happen.  Why not?  First, the periodic measurements of PSA on which to base decisions to stop or start therapy meant that patients often dropped more, and sometimes way more than the desired goal of 50\%, and vice-versa for when therapy was actually restarted.  Virtually all models indicate a poorer performance of adaptive therapy with imprecise switch points that overshoot the targeted switch values. The second reason is decisive and it concerns enlisting the needed density-dependence.  If the tumor burden and cancer cell populations sizes are well below carrying capacity, then both cell types may enjoy positive fitness in the absence of therapy~\cite{zhang2017integrating, strobl2020turnover, hansen2020cancer}. As the tumor cell populations grow, per capita growth rates decline for both cell types, but with asymmetric competition, the resistant cells will experience negative fitness even as the sensitive cells continue to grow.  This is a sweet-spot.  In this region of high tumor burden, the sensitive cells not only retard but reverse the growth rate of the resistant cells. Our retrospective analysis supports the conclusion of a number of mathematical and theoretical investigations.  Namely, adaptive therapy seems to work best if 1) the threshold for ceasing therapy is quite high (80\% rather than 50\%, for instance), and 2) the overall tumor burden is maintained as high as possible without endangering the patient~\cite{kim2021adaptive, hansen2020modifying, viossat2021theoretical}.  

Theory is asking the physician to hold off resuming therapy until as late as possible in terms of the recovery of the tumor during a period of no therapy.  With retrospective analysis of each patient, we find that failure of indefinite control happened because tumor burdens during periods of no therapy were being kept too low to permit sufficient density-dependence to facilitate negative growth rates of the resistant cells. For these patients, therapy was resumed too soon! Yet, there are likely dangers associated with containment strategies aiming for overall high, and persistent tumor burdens.  These fall into four important categories: 1) patients becoming symptomatic, 2) the ability of the biomarker to be sufficiently accurate and measurable frequently, 3) subsequent cancer evolution during the period of adaptive therapy, and 4) risk of new metastases.

An adaptive therapy clinical trial on metastatic thyroid cancer (NCT03630120) illustrates two of these concerns.  The trial was suspended.  The suspension was not mandated by the required stopping criteria, but because of two unanticipated issues.  In one of the patients receiving adaptive therapy there was an initially good response, therapy was stopped and the tumor burden allowed to recover.  However, the patient began to feel pain and other ill effects of the tumor burden before it had recovered to pre-treatment level.  The patient had become symptomatic at which point either treatment must be resumed or a different course of therapy considered.  In another patient, after the resumption of therapy the tumor continued to grow, at least based on the biomarker, with no indication of a future decline (Christine Chung and Joel S. Brown, unpublished data).  The small number of patients in both the randomized control arm (continuous therapy) and the adaptive therapy arm precluded meaningful interpretation of the potential efficacy of adaptive therapy.  But, it pointed to issues of patients becoming symptomatic, of the reliability of the biomarker as an accurate indicator of tumor burden, and of rapid changes in tumor burden and disease disposition that occurred at a faster time-scale than the ability to adjust therapies. 

While not yet documented in any clinical trial of adaptive therapy, there remains the concern that high tumor burden, subjected to on and off therapy cycles may incubate additional mutations and adaptations by the cancer cells.  For instance, with time, the resistant cancer cells may evolve traits that minimize or even reverse the cost of resistance. A large residual tumor population is more likely than a very small tumor burden to give rise to such mutations that propel the cancer cells to greater levels of malignancy.  As of yet, models of adaptive therapy have not considered the risk of additional progressive evolution beyond that expected regarding drug sensitivity.

Current trials and most models of adaptive therapy consider the patient’s total tumor burden even when it is known that the cancer is spread across several or many metastatic sites.  As the overall tumor burden shrinks different lesions may not necessarily respond similarly or proportionately. Furthermore, the fraction of resistant and sensitive cells may differ among lesions, particularly if they are in different tissues.  Over the course of adaptive therapy, the tumor burden may become more dispersed among lesions, more concentrated in a lower number of tumors, or, of most concern, metastasize to new sites.  In the case of melanoma, whether under standard of care continuous therapy or as an adaptive therapy trial (NCT03543969), there is a risk of the disease metastasizing to the brain even as therapy efficacy may be good elsewhere in the patient’s body. A recent spatial model of the prostate clinical trial, imagined dynamics both within and between metastatic sites in response to the on and off cycles of therapy.  The model predicts that the adaptive therapy regimen will vary if the disease is represented by many small versus a few large tumors~\cite{gallaher2022sum}.

As more clinical trials of adaptive therapy emerge from integrating mathematical models with clinical opportunity and need, it will be essential to consider the trade-offs associated with maintaining relatively large tumor burdens.  A large tumor burden may increase the efficiency of an adaptive therapy regimen while increasing the risks of additional progressive evolution, other ill effects of tumor burden, and the appearance of new lesions within the same or different tissues.  Balancing these costs and benefits will likely be disease and drug specific, and will require a continued lockstep between mathematical models and empirical studies, data and observations of patients.

\subsection{Can a mathematical model drive treatment decision making?}
\color{red}Mark Robertson-Tessi \& Sandy Anderson: \color{black}
Many models of adaptive therapy are currently hypothesis-generating models with less focus on predictive insight~\cite{enderling2020all}. One of the main goals of personalized therapy is the ability to predict likely tumor dynamics arising from all available treatment options, and then select the most promising. Therefore, there is a great need for clinically suitable predictive mathematical models that track patient-specific tumor dynamics. There are, however, numerous challenges that need to be surmounted for this approach to be broadly successful and able to be scaled to large numbers of patients.

\begin{figure}[t!]
    \centering
    \includegraphics[width=1\linewidth]{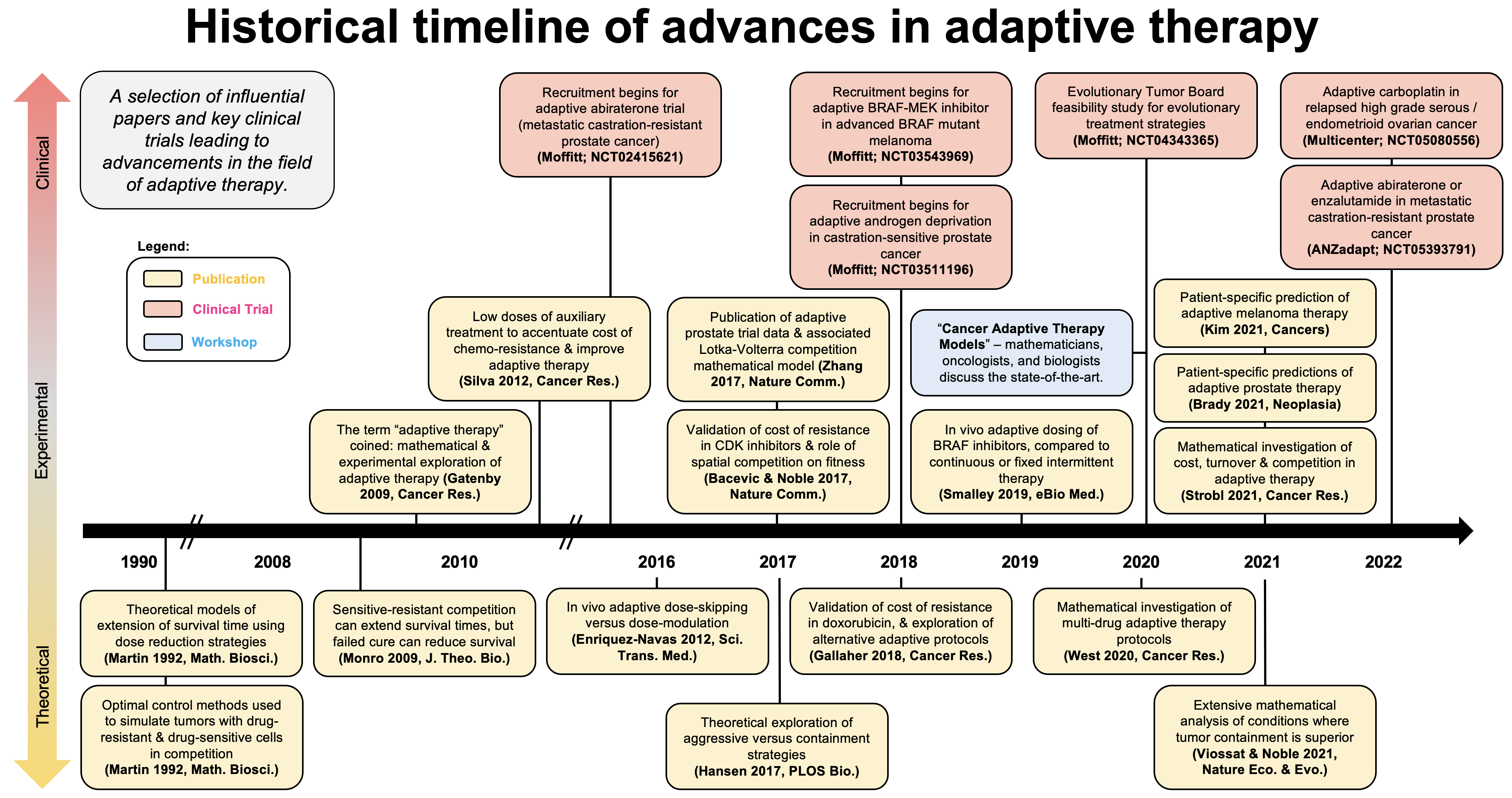}
    \caption{{\bf Timeline of advancements in adaptive therapy}: A selection of influential papers and key clinical trials leading to advancements in the field of adaptive therapy. This selection includes papers with experimental or clinical adaptive data in addition to well-cited theoretical publications.}
    \label{fig:adaptive_therapy_timeline}
\end{figure}

To begin with, clinical data are often sparse and often collected at times of little value to a mathematical model. For example, imaging scans are often collected some time prior to therapy, and then, depending on the cancer, the first follow-up scan occurs after 6 to 12 weeks. Yet, in many models of different diseases, the relevant action that would inform therapeutic decisions via modeling  occurs during the immediate aftermath of therapy. Furthermore, the models benefit greatly from multiple data points, so it may take several follow-up scans to narrow the parameterization of a model to determine treatment efficacy and tumor dynamics, by which time the tumor may have progressed and thus no prediction is needed. In a sense, current imaging schedules are designed to detect progression reasonably soon after it occurs, often using the standard RECIST criteria or similar, while balancing the costs of repeated imaging. To move into predictive mathematical modeling, we must rethink how we collect patient data, in that we don't want to simply detect progression, but rather quantify and understand the tumor dynamics prior to progression such that one can inform treatment decisions to avoid it.

A second consideration is the uncertainty of the entire system, from patient to model. Even an optimally designed data collection protocol that would optimize model predictability will still leave significant uncertainties in such predictions. The idea that we will have the data and modeling to precisely predict the future course of a patient's disease is not realistic, or even possible. Therefore, all predictions must include statistically rigorous uncertainty analyses. The concept of the `Phase i' trial~\cite{kim2016phase, scott2012phase} has been developed as a framework to quantify and apply such uncertainties. The basic principle is to develop cohorts of virtual patients derived from a calibrated mathematical model, and apply ``clinical trials'' to such cohorts with varying regimens of treatment. Much like changes to the standard of care that arise from cohort-to-cohort comparisons, the same approach is used in the virtual cohorts to decide on the treatment that is most likely to succeed. In general, the complete virtual cohort is formed by examining historical data on the disease, both at the individual patient level and via cohort outcomes from clinical trials. Once this global cohort is established from the model, an individual patient can be compared to each virtual patient, and matched to those that exhibit similar dynamics as the real patient. This patient-specific virtual subcohort is then subjected to the available therapy options, and their outcomes as a cohort are compared. The results provide decision-support for the treating physician. Using mechanistic models - as opposed to statistical models alone - has the advantage that as more patient data are collected on follow-up visits, the calibration and refinement of the patient's subcohort can be improved by removing virtual patients that responded differently.

Adaptive therapy strategies are one part of a broader approach to introduce evolutionary principles into dose scheduling to mitigate the evolution of resistance~\cite{gatenby2020integrating, noorbakhsh2020treating, belkhir2021darwinian, stackle2018}. Ongoing or planned evolution-based treatment trials include an extinction therapy trial in rhabdomyosarcoma (NCT04388839)~\cite{reed2020evolutionary}, adaptive androgen deprivation for castration-sensitive prostate cancer (NCT03511196), adaptive abiraterone or enzalutamide in castration-resistant prostate cancer (ANZadapt; NCT05393791), adaptive administration of BRAF-MEK inhibitors for Advanced BRAF mutant melanoma (NCT03543969), adaptive carboplatin in ovarian cancer (ACTOv trial), and a feasibility study for implementing evolution-based strategies with the aid of mathematical modeling decision-support (the ``evolutionary tumor board'' at the Moffitt Cancer Center; NCT04343365). Given the complexity of cancer as an evolutionary disease, many of these trials have been planned with insights gained from mathematical models. Figure \ref{fig:adaptive_therapy_timeline} shows a selection of influential papers and key clinical trials leading to advancements in the field of adaptive or evolution-based treatment. 

This figure illustrates that experimental and clinical investigation of adaptive therapy has progressed synergistically with mathematical and computational modelling. Perhaps the most important challenge concerns the communication between different disciplines: the practice of oncology, the theories of ecology and evolution, and the application of mathematical models to data were not historically in sync in terms of possibilities for both practice and outcome. An integrated team science approach focused on gaining a deeper understanding of each disease and the implications of each decision during treatment is key for the future success of an evolution-based cancer management.

\newpage
\printbibliography
\end{document}